# Comment

# 50 years of quantum spin liquids

Steven Kivelson & Shivaji Sondhi   Check for updates

In 1973, Philip Anderson published a paper introducing the resonating valence bond state, which can be recognized in retrospect as a topologically ordered phase of matter — one that cannot be classified in the conventional way according to its patterns of spontaneously broken symmetry. Steven Kivelson and Shivaji Sondhi reflect on the impact of this paper over the past 50 years.

Fifty years ago Philip W. Anderson[1] asked a deceptively simple question: might the ground state of a spin $S = 1/2$ quantum antiferromagnet on a triangular lattice fail to exhibit any form of broken symmetry? Most systems form broken symmetry states at low enough temperatures. Just as all liquids, with one exception, crystalize at low enough temperatures, $T < T_{freeze}$, the magnetic moments in insulating crystals typically form magnetically ordered ground states. For example, at temperature $T = 0$, if quantum fluctuations of the spin orientation are negligible, the moments on a triangular lattice form the three-sublattice antiferromagnetic order (Fig. 1a). In this situation, the directionality of the moments breaks the underlying spin-rotational symmetry of the problem, the different direction of the moments on different sublattice sites breaks the translational symmetry of the underlying lattice, and the existence of equilibrium moments at all means that time-reversal symmetry is spontaneously broken. Indeed, the primary 'Landau' framework for classifying distinct phases of matter is to identify them by their distinct patterns of spontaneously broken symmetries. The one element that remains a 'quantum liquid' as $T \to 0$ is helium. By analogy, a regular lattice of $S = 1/2$ moments in which quantum fluctuations are large enough that no magnetic order arises even at $T = 0$ has come to be called a 'quantum spin liquid'.

**A new quantum phase of matter**
In his paper, Anderson introduced a novel state — that he christened a "resonating valence bond" (RVB) state — which he proposed might capture the essence of the true ground state of the $S = 1/2$ antiferromagnetic Heisenberg model on a triangular lattice. It is now well established that, instead, the ground state for this particular problem has the same pattern of broken symmetry shown in Fig. 1a as that of the corresponding classical ($S \to \infty$) antiferromagnet. Nevertheless, the ideas Anderson introduced in his 1973 paper have turned out to be extremely influential and are now a part of the mental furniture of all modern quantum condensed-matter physicists. Indeed, from a contemporary perspective, the RVB paper can be recognized as being one of the first salvos of a 'post-Landau' era of condensed-matter physics — in which the classification of phases of matter was expanded from Landau era considerations of symmetry breaking alone. Other key papers from this time include the nearly contemporaneous Kosterlitz–Thouless[2] work on topological phase transitions and Wegner's slightly earlier construction of (what slightly later were identified as) Ising gauge theories in which no local order parameter distinguishes the phases[3].

In attempting to write an explicit wave-function for a state in which quantum fluctuations preclude any magnetic order, Anderson faced a significant technical hurdle. In order that the state preserve spin rotation symmetry, he chose to work in a physical basis in which pairs of spins form singlets that, following Pauling, he called valence bonds (Fig. 1b). His next suggestion was what set the paper apart — that the ground state would be a coherent superposition of possible ways of pairing up nearby spins into valence bonds, hence the RVB state.

**14 years later**
Initially, it was not clear that revolutionary implications followed from these details. Indeed, for the next 14 years the paper attracted only 70 citations until Anderson himself came back to the idea in 1987[4], following the discovery of cuprate superconductivity. At this point his original ideas combined with a beautiful set of developments on fractionalized excitations in systems such as polyacetylene and the fractional quantum Hall effect to set in motion a remarkable wave, yet unabated, of developments now typically referred to as the study of topologically ordered systems.

The existence of fractionalized excitations in 1D systems had been known for some time, including explicit notions of spin–charge separation introduced by Luther and Emery in the mid 1970s[5]. However, the relevance of these ideas to measurable systems, their connection to topological excitations, and their visualization in terms of patterns of valence bond order — all of which contributed to the post 1987 developments in the field of spin liquids (not to mention topological insulators) — was initiated by the work of Su, Schrieffer and Heeger[6] in the context of solitons in polyacetylene. The existence of a 2D fluid with fractionalized excitations was first realized by Laughlin[7] in the context of the theory of the fractional quantum Hall effect. (The modern perspective on these developments, and their relation to the physics of spin liquids and gauge theories has been recently reviewed in ref. [8].)

In the 1987 paper, Anderson suggested that one should think of valence bonds as pre-formed Cooper pairs, which become liberated to conduct upon light doping of the parent insulating state. Much like Anderson's first paper, the explicit propositions of the second have not stood the test of time, either concerning the nature of the insulating state in the cuprates nor in providing a generally accepted framework for understanding the superconducting state.

However, in common with Anderson's earlier paper, the ideas in his 1987 paper inspired an entire field and generation of physicists to action and discovery.

**Far-reaching implications**
Indeed, today we would describe the RVB state as the simplest example of a topologically ordered 'spin liquid', one with $Z_2$ topological order to be precise. The state does not break any symmetry generated by local operators. Instead its non-triviality shows up in four ways. One, there is a ground state degeneracy that depends on the topology of the underlying manifold. Two, there exists a fractionalized 'matter' excitation, the $S = 1/2$ charge neutral 'spinon'. Three, there is an emergent 'gauge' excitation, the 'vison', which has non-trivial braiding with the spinon. Four, all of this physics is encoded into a topological quantum field theory (TQFT). Interestingly $Z_2$ topological order is also what characterizes the deconfined phase of the Ising gauge theory — so in retrospect both Anderson and Wegner were introducing the same physics at about the same time!

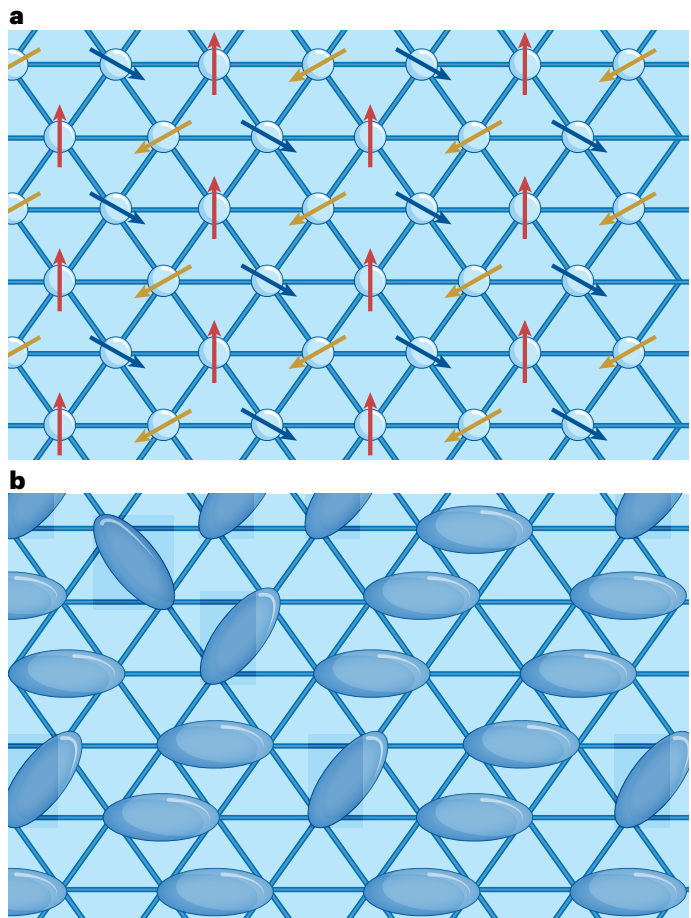

**Fig. 1 | Schematic representation of spins on a triangular lattice. a**, A three-sublattice antiferromagnetically ordered state. **b**, A typical valence-bond state.

It should be noted that much of the understanding of this cluster of ideas was, in fact, first worked out for the case of the fractional quantum Hall effect. Here the corresponding TQFTs are the celebrated Chern–Simons theories, and, moreover, essential features of the fractionalized phases have been unambiguously observed experimentally.

Since then these ideas have been expanded to cover a bewildering variety of theoretically well-characterized phases that involve emergent gauge fields of various kinds and quasi-particle excitations that generalize the visons and spinons from above.

The study of topological order in many-body physics has proven extremely fruitful. It has crystallized a new perspective on gauge theories in which the gauge structure is emergent rather than fundamental[9]. It has also inspired proposed approaches to topological quantum computing and error correction. On the theory end there is now a large set of solvable models with stable ground-state phases exhibiting various forms of topological order. The need of the hour is for experimental systems beyond the quantum Hall system — either in materials or in artificial platforms — that unambiguously realize more of these predicted phases. On the upside, an increasingly large number of materials — known as 'spin liquid candidate materials' — have been identified that do not magnetically order down to the lowest accessible temperatures and which more generally exhibit properties that are difficult to understand as reflecting the properties of any obvious candidate phase without topological order[10]. On the downside, in no cases is there unambiguous and direct evidence (for instance clear observation of fractionalized excitations) that some conventional but complicated explanation — for example, involving sample imperfections of one sort or another — can be definitively excluded.

Steven Kivelson[1,2] ✉ & Shivaji Sondhi[2]
[1]Stanford Institute for Theoretical Physics, Stanford University, Stanford, CA, USA. [2]Rudolf Peierls Centre for Theoretical Physics, University of Oxford, Oxford, UK.
✉e-mail: kivelson@stanford.edu

**Competing interests**
The authors declare no competing interests.